\documentstyle[11pt,newpasp,twoside]{article}
\begin{document}
\title{The Multiple Stellar Populations in $\omega$~Centauri}
\author{E. Pancino}
 \affil{Dipartimento di Astronomia, Universit\`a di Bologna, Via
  Ranzani 1, I-40127 Bologna, Italy}
 \affil{European Southern Observatory, K. Schwarzschildstr. 2,
  D-85748 Garching bei M\"unchen, Germany}


\begin{abstract}
In this contribution I am going to present some preliminary results of
a high-resolution spectroscopic campaign focussed on the most metal
rich red giant stars in $\omega$~Cen. This study is part of a long
term project we started a few years ago, which is aimed at studying
the properties of the different stellar populations in
$\omega$~Cen. The final goal of the whole project is the global
understanding of both the star formation and the chemical evolution
history of this complex stellar system.
\end{abstract}


\section{The Discovery of the RGB-a}

\begin{figure}
\plotfiddle{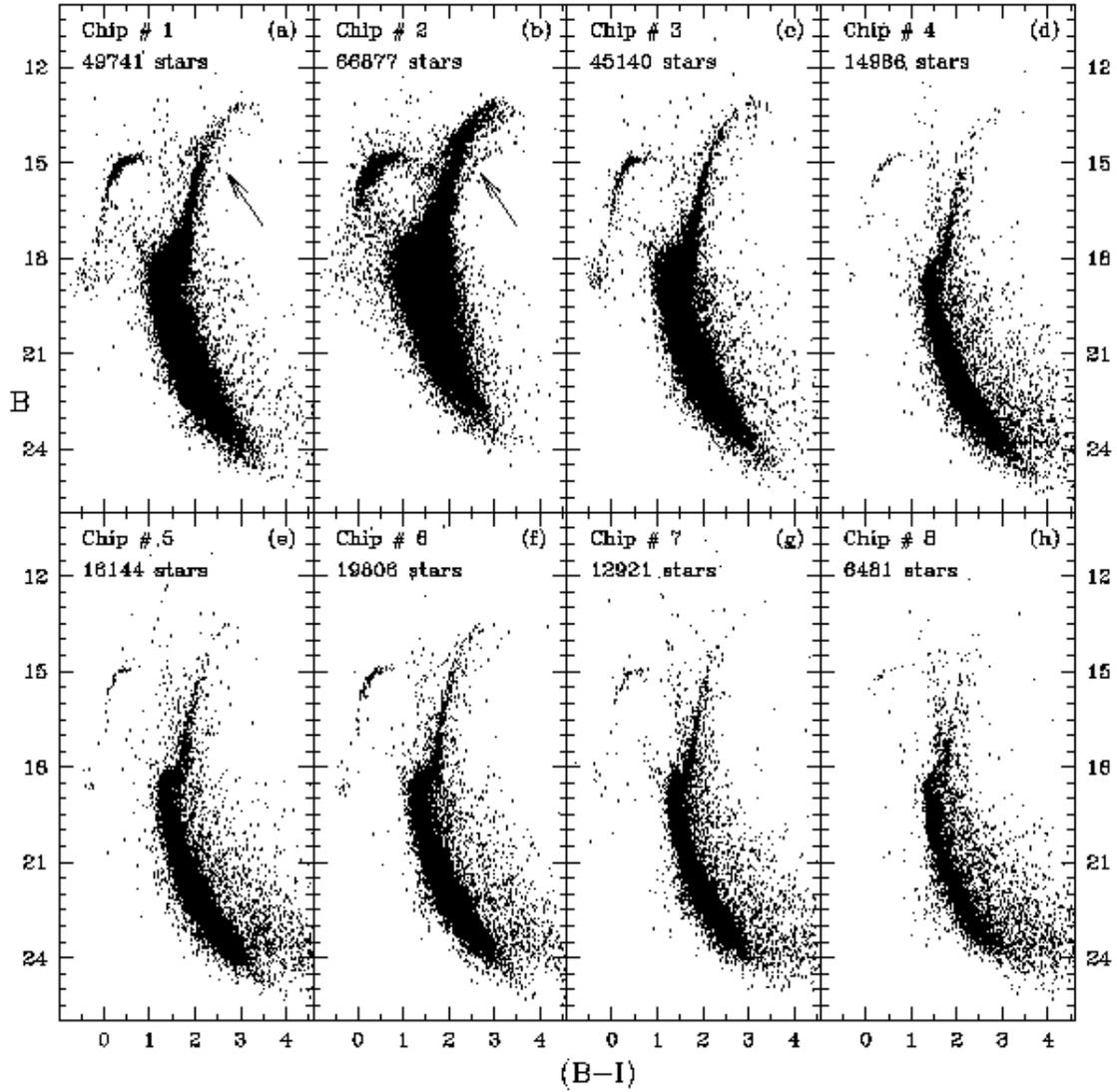}{15cm}{0}{70}{70}{-210}{-70}
\caption{The WFI photometry is displayed. Each panel shows the CMD
for one of the CCD chips in the WFI mosaic. The cluster center is
roughly located in chip \#~2. The arrows mark the RGB-a.}
\end{figure}

\begin{figure}
\plotfiddle{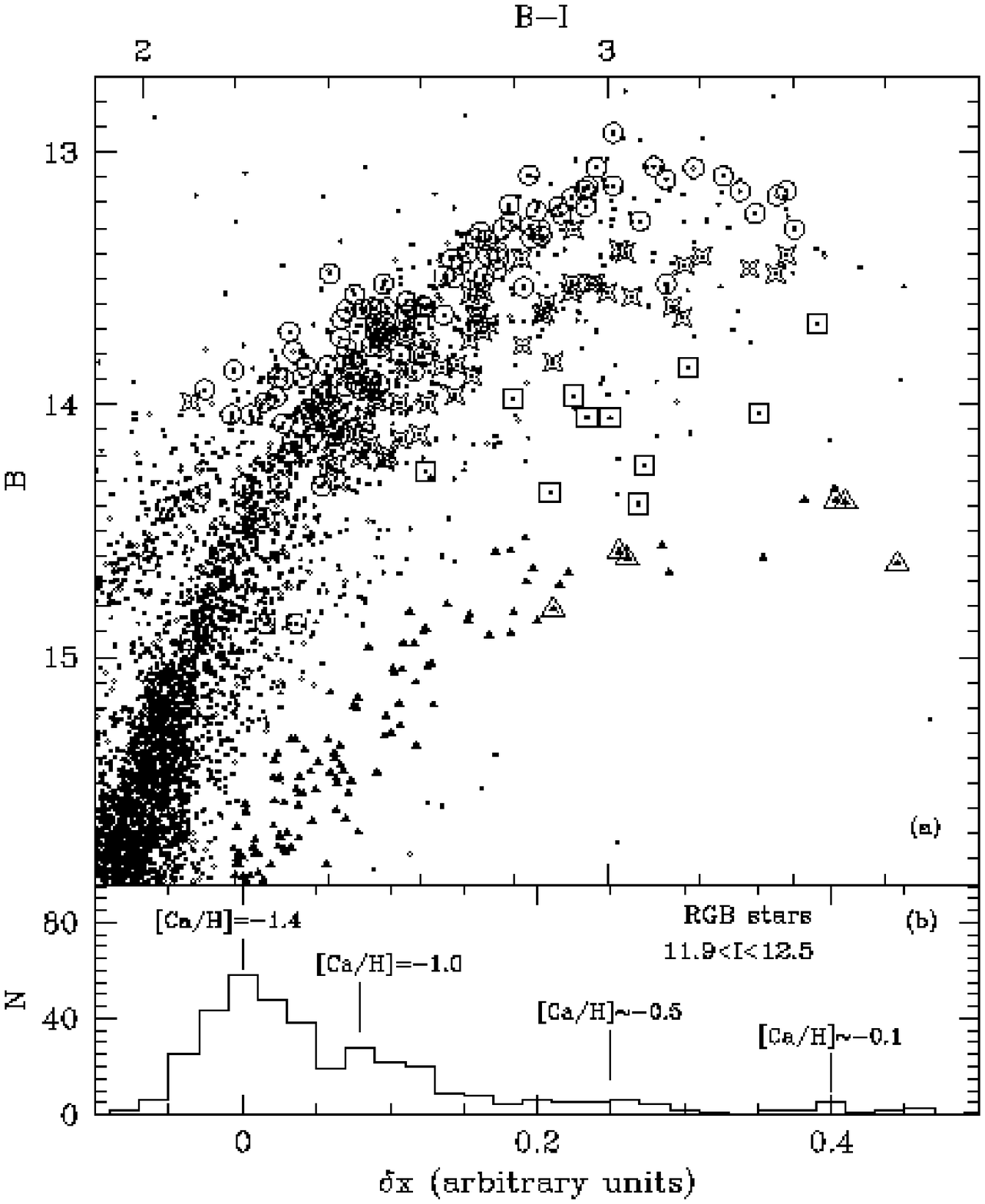}{14cm}{0}{70}{70}{-220}{-80}
\caption{{\it Top Panel:} Zoomed CMD of the upper RGB region. Stars in the
RGB-a have been plotted as {\it small filled triangles}. Large empty
symbols show the position in the CMD for stars with known metallicity
(from Norris et al. 1996). Different symbols refer to stars with
different metallicities: {\it large open circles} for stars with
$-1.5<$[Ca/H]$<-1.3$, {\it large open stars} for $-1.1<$[Ca/H]$<-0.85$,
{\it large open squares} for $-0.65<$[Ca/H]$<-0.4$, {\it large open
triangles} for [Ca/H]$>-0.3$, respectively. {\it Bottom Panel:}
Histogram of the distribution of the distances from the mean ridge
line of the dominant, metal poor RGB. The mean [Ca/H] abundances for
the four main components of the RGB are also marked.}
\end{figure}

The first surprising result that we got in the study of $\omega$~Cen
came from the wide field photometry that we performed in 1999 at the
2.2 m ESO-MPI telescope (La Silla, Chile), equipped with the WFI (Wide
Field Imager) mosaic camera.

Figure~1 shows the $(B,B-I)$ CMDs for each of the 8 WFI chips
separately. The cluster is roughly centered on chip \#2.  More than
230,000 stars have been plotted: to our knowledge, this is the largest
stellar sample ever observed in $\omega$~Cen. The most striking
feature of this CMD is the existence of a {\it complex} structure in
the brighter part of the Red Giant Branch (RGB): at least two main
components are visible. Particularly notable is the presence of a
narrow sequence (see also Lee et al. 1999), significantly redder and
more bent than the bulk of the ``main'' RGB stars, which we call the
{\it anomalous} RGB (hereafter RGB-a). Note that the RGB-a is visible
only in the CMD from chip \#2 (where the cluster center is located)
and to a much lesser extent in the nearby chip \#1 and possibly chip
\#3.

The morphology of the RGB-a and its position in the CMD strongly
suggest that it is populated by stars much more metal rich than the
$\omega$~Cen bulk population. In order to get a first indication about
the metal content of the RGB-a stars, we have combined our catalog
with the extensive low resolution spectroscopic survey carried on by
Norris et al. (1996), based on the infrared calcium triplet indicator:
Figure~2 shows the metallicity distribution of the red giants in
$\omega$~Cen, obtained by using the distance of each star from the
mean ridge line of the dominant, metal poor population as a sort of
metallicity indicator (Figure~2, Panel {\it (b)}). We can thus identify
three different RGB components, or sub-populations:

\begin{itemize}
\item{{\it RGB-MP:} the dominant, metal poor population, with
[Ca/H]$\sim-1.4$, which drives the average metallicity of the whole
cluster.}
\item{{\it RGB-MInt:} the intermediate metallicity population, with
[Ca/H]$\sim-1.0$ and a long extended tail reaching up to
[Ca/H]$\sim-0.5$.}
\item{{\it RGB-a:} the anomalous, newly discovered population, with
[Ca/H]$\sim-0.1$, which represents the extreme metallicity end of the
cluster distribution, and comprises $\sim5\%$ of the whole RGB
population.}
\end{itemize}

Thus, we have photometrically isolated a distinct sub-population, the
RGB-a, whose sharp and clear-cut shape in our CMD allows us to select
a representative sample of stars for high-resolution spectroscopic
follow up, in order to obtain their detailed abundance pattern.

\section{The UVES Spectroscopic Campaign}

\begin{figure}
\plotfiddle{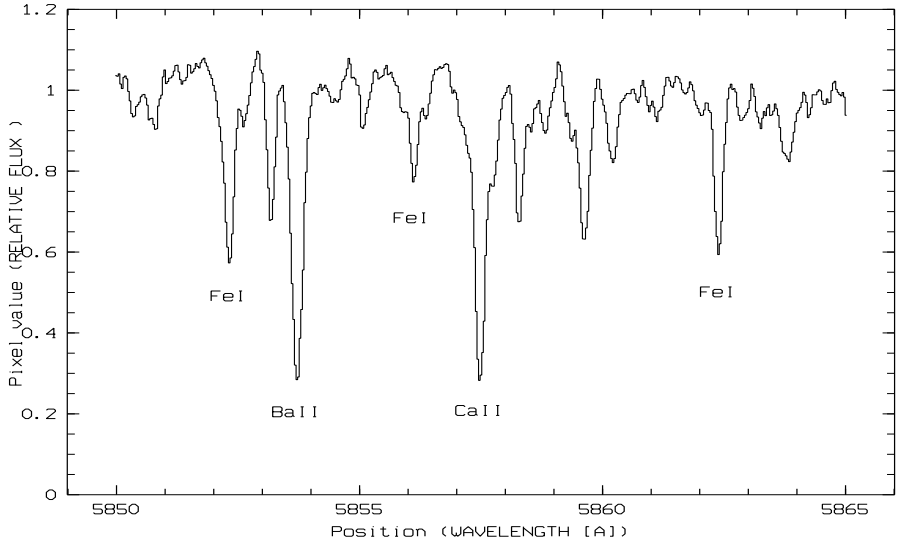}{7.5cm}{270}{50}{40}{-220}{220}
\caption{A sample of the spectra obtained during the UVES pilot
observing run in june 2000 ({\it Run A}). The signal to noise ratio of
these spectra is $S/N\sim100-150$.}
\end{figure}

\begin{figure}
\plotfiddle{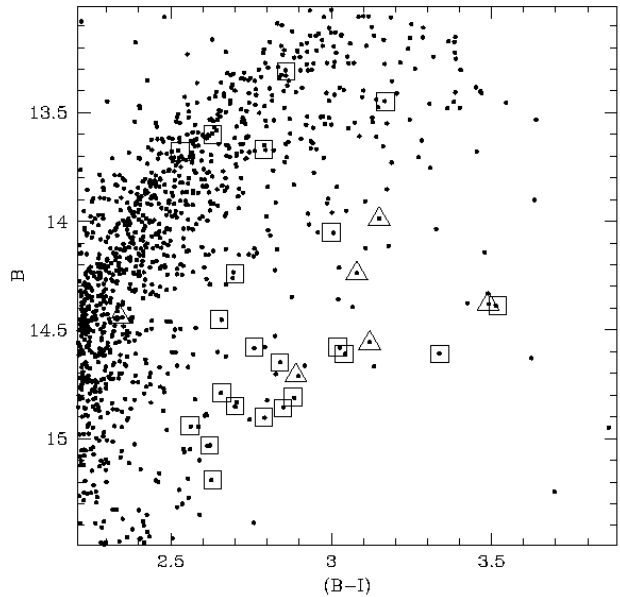}{8cm}{0}{40}{40}{-140}{-50}
\caption{The targets for the two UVES runs are marked on the upper
part of the RGB. Large open triangles show the position of the 6 {\it
Run A} targets while large open squares refer to the 23 {\it Run B}
targets.}
\end{figure}

The discovery of the RGB-a poses a number of questions that need to be 
answered, in the light of what we already know about the chemical and
dynamical properties of the RGB stars in $\omega$~Cen, to put this new 
stellar population in the context of the star formation and/or merging 
history of the whole cluster.

In order to completely characterize the chemical properties of the
RGB-a stars, we need to obtain high resolution, high signal-to-noise
spectra for a sufficient number of stars. The ideal tool is UVES
(UV-Visual Echelle Spectrograph) mounted on the VLT (Very Large
Telescope) at Paranal, in Chile. With this instrument we can obtain a
resolution of $R\sim$45000, comparable to the latest studies (like
Smith et al. 2000), and we can easily and fastly achieve
S/N$\sim$100--150.

Our observing programme was split in two different observing runs: a
pilot study on six RGB stars, that was carried out in june 2000 ({\it
Run A}) and a more extensive survey of 23 additional stars, that was
successfully carried out in april--may 2001 ({\it Run B}). The main
goals of this project are the following: {\it (a)} to sample
adequately the whole extension of the RGB-a and {\it (b)} to observe a
few RGB-MP and RGB-MInt stars in common with previous high resolution
studies, both to have a feeling for the soundness of our results and,
lately, to merge all the existing databases in a homogeneous
way. Figure~4 shows the position of the programme stars along the RGB.

In the following sections I will briefly describe the method used and
report on some preliminary results for the six stars of the pilot
survey, while the complete analysis of the whole sample (29 stars in
total) is still under way.

\section{Observational Material}

We selected our programme stars for {\it Run A} among the
metal-richest stars in $\omega$~Cen (see Table~1). Three of them,
stars ROA~300, WFI~222068 and WFI~222679, belong to the anomalous
RGB-a, while the other three belong to the intermediate metallicity
population, the RGB-MInt.

None of the six stars has been observed before with high-resolution
spectroscopy ($R\geq20000$), except for the RGB-MInt star ROA~371 (see
Table~2), that has been studied by Paltoglou \& Norris (1989), Brown
et al. (1991), Vanture et al. (1994) and Norris \& Da Costa (1995),
and for this reason is the ideal comparison object. ROA~371 has always
been considered to be a peculiar star, showing strong BaII and SrII
lines (Dickens \& Bell 1976), and a mild variability (Cannon \& Stobie
1973).

\begin{table}
\begin{center}
\begin{tabular}{lcccccccl}
\hline
Star & $V$ & $B$--$V$ & $T_{eff}$ & $T_{ex}$ & $\log g$ &
$v_t$ & [Fe/H] & Pop \\
\hline
ROA~300    & 12.71 & 1.55 & 3900 & 3900 & 0.9 & 1.3 & -0.78 & RGB-a \\
WFI~222068 & 12.95 & 1.49 & 3950 & 4000 & 1.1 & 1.3 & -0.49 & RGB-a \\
WFI~222679 & 13.26 & 1.33 & 4150 & 4100 & 1.2 & 1.4 & -0.54 & RGB-a \\
ROA~211    & 12.43 & 1.44 & 4000 & 4000 & 0.8 & 1.9 & -1.02 & RGB-MInt \\
ROA~371    & 12.71 & 1.41 & 4050 & 4000 & 0.7 & 1.5 & -0.95 & RGB-MInt \\
WFI~618854 & 13.26 & 1.06 & 4500 & 4600 & 1.2 & 1.5 & -1.20 & RGB-MInt \\
\hline
\end{tabular}
\end{center}
\caption{Programme Stars Parameters. The columns contain the following 
information: {\it (1)} the star name from Woolley (1966) or from the
WFI photometry of Pancino et al. (2000); {\it (2)-(3)} magnitudes and
colors from Pancino et al. (2000) -- the unpublished $V$ magnitude
comes from the same dataset; {\it (4)} the photometric estimate of
$T_{eff}$ obtained with the calibration by Alonso et al. (1999); {\it
(5)-(8)} the stellar parameters resulting from our abundance analysis
(see text) and {\it (9)} the sub-population to which the star belongs,
according to Pancino et al. (2000).}
\end{table}

\section{Abundance Analysis}

{\it Run A} was successfully completed in june 2000, by obtaining
high-resolution ($R\sim 45000$) echelle spectra with S/N ranging from
$\sim 100$ to $\sim 150$ (an example of the quality of these spectra
is displayed in Figure~3). The monodimensional spectra were extracted
using the UVES ESO pipeline, then continuum normalized and corrected
for telluric absorption with IRAF\footnote{IRAF is distributed by the
National Optical Astonomy Observatories, which is operated by the
association of Universities for Research in Astronomy, Inc., under
contract with the National Science Foundation.}. All of the six stars
turned out to be members of $\omega$~Cen on the basis of the radial
velocity derived from our spectra.

A first estimate of the $T_{eff}$ and $\log g$ has been derived from
absolute magnitudes and dereddened colors of our programme stars (see
Table~1), obtained from Pancino et al. (2000), by assuming
$E(B-V)=0.12$ and $(m-M)_V=13.92$, according to Harris (1996) and by
using the temperature scale calibration and bolometric corrections
from Alonso et al. (1999).

The abundance analysis was done on a set of reliable and unblended
spectral absorpion lines that span a range in strength, excitation
potential and wavelength. The results presented here are based on 94
FeI lines, 10 FeII lines, 17 CaI lines and 10 SiI lines in the
wavelength range $\sim 5300 - 6800$\AA. The atomic data and $\log gf$
are taken from the available literature among the most recent
laboratory measures: we used mainly the NIST\footnote{NIST (National
Institute of Standards and Technology) Atomic Spectra Database,
Version 2.0 (March 22, 1999), \tt
{http://physics.nist.gov/cgi-bin/AtData/main\_asd}.} database and Nave
et al. (1994) for iron. Equivalent widths were measured with IRAF, by
gaussian fitting of the line profile on the local continuum. The
gaussian profile is a good approximation for these low-gravity, cool
stars, if one avoids strong lines. For the coolest stars we chose not
to measure any atomic line inside the prominent TiO bands.

The abundance calculations were made using an extension of the OSMARCS
grid of flux-constant, plane parallel, LTE model atmospheres for M
giants and supergiants calculated by Plez (1992), in the [Fe/H]$=-1.0$
to $-0.3~dex$ range (Plez 1995, private communication). First
estimates of the microturbulent velocity $v_t$ (and of [Fe/H]) were
derived from curves of growth for FeI and FeII.

We then refined our parameters set by enforcing simultaneously the
following conditions: {\it (a)} $v_t$ was determined by imposing that
strong and weak FeI lines give the same abundance; {\it (b)} the
temperature was constrained by imposing excitation equilibrium for
FeI; {\it (c)} the surface gravity was determined by imposing
ionization equilibrium between FeI and FeII. The resulting best
parameters for the six programme stars are shown in Table~1:
temperatures and gravities resulting from our analysis are close to
the photometric values, and we will adopt them for the rest of our
treatment.

\begin{table}
\begin{center}
\begin{tabular}{lrrrrr}
\hline 
& PN89 & B91 & V94 & ND95 & {\it Here} \\
\hline
$R$       &$\sim17000$&$\sim17000$&$\sim20000$&$\sim38000$&$\sim45000$\\
$S/N$     &$\leq 50$  &$\sim 100$ & 70--150   &$\sim 50$  & 140       \\
$T_{eff}$ & 4000      & 4000      & 4000      & 4000      & 4000      \\
$\log g$  & 0.9       & 0.9       & 0.9       & 0.9       & 0.7       \\
$v_t$     & 2.5       & 1.5       & 2.2       & 1.6       & 1.5       \\
$$[Fe/H]  & -1.37     & -0.9      & -1.00     & -0.79     & -0.95     \\
\hline
\end{tabular}
\end{center}
\caption{Comparison with previous literature results for star
ROA~371. Resolution, signal-to-noise ratio, stellar parameters and
derived iron abundances are shown for: {\it (PN89)} Paltoglou \&
Norris (1989); {\it (B91)} Brown et al. (1991); {\it (V94)} Vanture et
al. (1994); {\it (ND95)} Norris \& Da Costa (1995) and {\it (Here)}
this contribution.}
\end{table}

The results we obtained can be directly compared with previous
studies, since star ROA~371 has been observed before by four different
groups, with the results reported in Table~2. The stellar parameters
of the different studies are in good agreement with each other, and
the [Fe/H] abundance determinations are consistent within the
uncertainties.

Moreover, we observed star ROA~300 with low and intermediate
resolution spectroscopy in the infrared wavelength range with SOFI
(see the contribution by Ferraro, Pancino \& Bellazzini, this volume)
and we obtained a preliminary abundance of [Fe/H]$\sim$-0.9, again in
good agreement, within the uncertainties, with the value obtained from 
the UVES spectra.

\section{Preliminary Results}

\begin{figure}
\plotfiddle{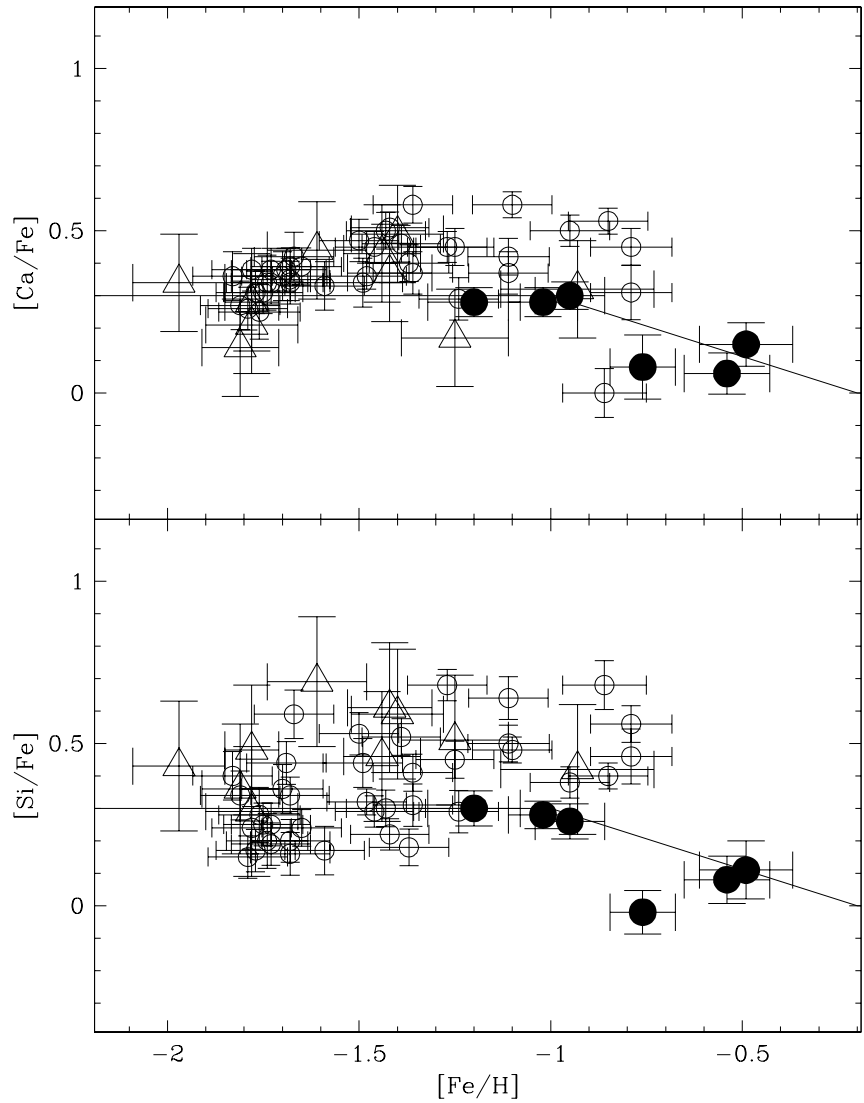}{14cm}{0}{60}{60}{-180}{-20}
\caption{The $\alpha$-enhancement of the giants in $\omega$~Cen is
explored through the element ratios [Ca/Fe] ({\it Top Panel}) and
[Si/Fe] ({\it Bottom Panel}). The abundances derived for the 6 stars
in {\it Run A} (solid dots) are compared with the studies of Norris \&
Da Costa (1995) (empty circles) and Smith et al. (2000) (empty
triangles).}
\end{figure}

The values listed in Table~1 are the first direct estimate of the
chemical abundance for stars belonging to the RGB-a population ever
obtained using high-resolution spectroscopy. By taking a straight
average for the three RGB-a stars we find: [Fe/H]=$-0.60\pm 0.16$ and
[Ca/H]=$-0.50\pm 0.17$.  This result confirms the conclusion by
Pancino et al. (2000) that the RGB-a is the extreme metal rich
component of the $\omega$~Cen stellar mix.  It is interesting to note
that the abundances obtained here are lower than previous estimates
based on calcium triplet measures (see above). However, it should be
noted that calcium triplet calibrations are usually quite uncertain at
the high metallicity end: Norris et al. (1996) explicitly warned the
reader on the fact that their metal-rich stars results are based on an
{\em extrapolation} of their calibration and should thus be treated
with caution.
 
However, the most interesting result that we got from the analysis of
the six spectra presented here is the different $\alpha$-enhancement
measured for the two sub-populations. We found [Ca/Fe]=$+0.28\pm 0.02$
and [Si/Fe]=$+0.28\pm 0.02$ for the three RGB-MInt stars, in perfect
agreement with the expected $\alpha$-enhancement for globular cluster
stars, and only [Ca/Fe]=$+0.10\pm 0.05$ and [Si/Fe]=$+0.12\pm 0.05$
for the three RGB-a stars in our sample. This is the first evidence
that stars with significant low $\alpha-$enhancement do exist in
$\omega$~Cen. Note that both Norris et al. (1996) and Smith et
al. (2000), who analyzed red giants in $\omega$~Cen with metallicities
up to [Fe/H]$=-0.78$ and [Fe/H]$=-0.95$ respectively, found {\em no
evidence} of decreasing Calcium (or any other $\alpha$-element)
enhancement with increasing metallicity.

The effect is also illustrated in Figure~5, where their studies (empty
symbols) are compared with our results (filled symbols).  Figure~5
clearly shows the tendency of decreasing [Ca/Fe] and [Si/Fe] with
increasing metallicity for the 6 stars presented here.  The solid line
marks the Galactic [$\alpha$/Fe] relation as can be desumed from
Edvardsson et al. (1993) and Gratton (1999).

If the result shown in Figure~5 is confirmed by larger samples of
metal rich stars, {\em we have found the first indication that the
medium from which the RGB-a stars have formed has been previously
enriched by type Ia Supernovae ejecta}.

\section{Implications for the formation scenarios in $\omega$~Cen}

A detailed discussion of the chemical evolution of $\omega$~Cen is
beyond the purpose of this contribution, however some obvious
implications of the results presented above deserve a short comment.
How do the new results presented here on the RGB-a population compare
with the most widely accepted scenarios for the formation of
$\omega$~Cen?

There are two main scenarios that have been proposed to explain the
variety of stellar populations observed in $\omega$~Cen, both based
upon non-negligible observational evidences: {\it Case a)} the most
popular scenario proposes that $\omega$~Cen is a complex stellar system
(possibly a dwarf galaxy, or the remnant of a formerly larger galaxy;
see Freeman 1993) that was able to self-enrich during its star
formation history; {\it Case b)} it has also been suggested (Icke \&
Alcaino 1988; Norris et al. 1997; Pancino et al. 2000) that some of
the present stellar components of the cluster may have formerly
belonged to an external, smaller stellar system that merged with
the main body of $\omega$~Cen. Let's discuss the two separately.

\vspace{0.3cm} 
{\it Case a)} Previous abundance studies on the RGB-MP and RGB-MInt
stars in $\omega$~Cen, like the ones by Norris et al (1995) and Smith
et al. (2000), show that: {\it (i)} the $\alpha$ enhancement is
constant ([$\alpha$/Fe]$\sim$+0.3) with metallicity, implying that
part of the SNe~II ejecta have been retained by the cluster; {\it (ii)}
there is no evidence for a contribution from SNe~Ia for any of the
stars studied so far ([Fe/H]$\la$-0.8) and {\it (iii)} the overabundance
of $s$-process elements is increasing with metallicity, probably
because the rich (and presumably younger stars) have been enriched by
the slow winds of AGB intermediate mass stars.

While the [$\alpha $/Fe] overabundance indicates that a major
interstellar medium enrichment by type II supernovae took place over a
relative short timescale ($\le$ 1 Gyr), the $s$-process element
overabundance indicates that the enrichment by AGB stars took place
over timescales ranging from a few hundred million and a few billion
years, depending on their actual progenitor mass. 

Moreover, the metal poor and intermediate populations show {\it no
sign of enrichment by type Ia supernovae ejecta}. This could be
explained both with a star formation process short enough to end
before the onset of type Ia supernovae, or, in case of longer star
formation timescales, by assuming that type Ia supernovae winds
efficiently removed most of their own products (Smith et
al. 2000). 

Hence, if we assume that indeed the detailed chemical composition of
{\em all} of the giants in $\omega$~Cen can be explained by pure
self-enrichment, then the evidence we presented above tells us that
the most metal rich (and thus somewhat younger) population of
$\omega$~Cen has been enriched by supernovae type Ia ejecta. In this
scenario we have detected the ``knee'' of the [$\alpha$/Fe] relation,
a very valuable constraint to the chemical evolution of the whole
stellar system (McWilliam 1997). The canonical interpretation of this
feature in the [$\alpha$/Fe] relation is that all the stars poorer
than the knee should have formed in a relatively short timescale of
the order of $\sim$1~Gyr. However, these timescales are still
uncertain, see for example Matteucci \& Recchi (2001) for a review of
the sensitivity of the SNIa enrichement timescale on initial mass
function and star formation rate.

Thus, within the framework of pure self-enrichment chemical evolution,
our finding would also imply some age spread ($\sim $few Gyr) between
the RGB-a population and the previous, more metal poor ones. An age
spread of $3-5~Gyr$ seems infact needed to explain the morphology of
the SGB-TO region of recent Str\"omgren photometries like the ones by
Hughes \& Wallerstein (2000) or Hilker \& Richtler (2000). However,
a clear answer to the relative ages problem can only come from
high-resolution spectroscopic studies of stars in the SGB-TO of
$\omega$~Cen, that can break the age-metallicity degeneracy.

\vspace{0.3cm} 
{\it Case b)} The merging hypothesis for the formation of $\omega$~Cen
was put forward by Icke \& Alcaino (1988), to explain both the
ellipticity and the metallicity spread observed.  Admittely, the
collected observational evidence make the merger of two ordinary
globular cluster in the potential well of the Milky Way unlikely. For
istance, accordingly with Smith et al. (2000) the merging of two or
more single metallicity clusters scenario cannot account for the broad
distribution of the RGB. However, evidence of correlation between the
metallicity and the dynamical properties (Norris et al. 1997) or the
spatial distribution (Pancino et al. 2000) are still difficult to
explain in the pure self-enrichment scenario.

Thus, the possibility still exists that at least the RGB-a stellar
population was originally a satellite system of $\omega$~Cen. Within
this framework the most promising scenarios seems to be the so-called
{\em merger within a fragment} (Norris et al. 1997), in which the
RGB-a stellar population was originally a satellite gas cloud in the
potential well of the larger {\it $\omega$~Cen galaxy}. The results
presented here add the evidence that this cloud was also previously
enriched by the ejecta of type Ia supernovae.

\section{Conclusions and Future Prospects}

The results presented here represent a first step towards the
understanding of the star formation history of $\omega$~Cen. A
comprehensive scenario of course requires a much more extensive sample
of stars, together with accurate measures of the relative ages and
detailed abundance patterns of stars near the Turn-off and Sub Giant
Branch regions of the CMD. We are already planning to extend our study 
in this direction within the framework of the global project presented 
by Ferraro, Pancino \& Bellazzini, in this volume.



\begin{references}
\reference Alonso, A., Arribas, S., Mart\'\i nez-Roger, C. 1999, A\&A, 140, 261
\reference Brown, J.A., Wallerstein, G., Cunha, K., Smith, V.V., 1991
A\&A, 249, L13 
\reference Cannon, R.D., Stobie, R.S. 1973, MNRAS, 162, 207
\reference Dickens, R.J., Bell, R.A. 1976, ApJ, 207, 506
\reference Edvardsson, B., Andersen, J., Gustafsson, B., Lambert,
D.L. Nissen, P.E. \& Tomkin, J. 1993, A\&A, 275, 101
\reference Freeman, K.C. 1993, in ASP Conf. Ser. 48, The Globular
Cluster--Galaxy Connection, ed. J.H. Smith \& J.P. Brodie (San
Francisco:ASP), 608 
\reference Gratton, R.G. 1999, Ap\&SS, 265, 157 
\reference Harris, W.E. 1996, AJ, 112, 1487
\reference Hilker, M. \& Richtler, T. 2000, A\&A, 362, 895
\reference Hughes, J. \& Wallerstein, G. 2000, AJ, 119, 1225
\reference Icke, V. \& Alcaino, G. 1988, A\&A, 104, 115
\reference Lee, Y-W., Joo, J-M., Sohn, Y-J., Rey, S-C., Lee, H-c., \&
Walker, A.R., 1999, Nature, 402, 55
\reference Matteucci, F. \& Recchi, S. 2001, ApJ, 558, 351
\reference McWilliam, A. 1997, ARA\&A, 35, 503
\reference Nave, G., Johansson, S., Learner, R.C.M., Thorne, A.P.,
Brault, J.W. 1994 ApJS, 94, 221
\reference Norris, J.E., Da Costa, G.S. 1995, ApJ, 447, 680 
\reference Norris, J.E., Freeman, K.C. \& Mighell, K.L., 1996, ApJ, 462, 241
\reference Norris, J.E., Freeman, K.C., Mayor, M. \& Seitzer, P. 1997,
ApJL, 487, L187 
\reference Paltoglou, G., Norris, J.E. 1989, ApJ, 336, 185 
\reference Pancino, E., Ferraro, F.R., Bellazzini, M., Piotto, G. \&
Zoccali M. 2000, ApJ, 534, L83
\reference Plez, B. 1992, A\&AS, 94, 527
\reference Smith, V.V., Suntzeff, N.B., Cunha, K., Gallino, R., Busso,
M., Lambert, D.L., Straniero, O. 2000, AJ, 119, 1239
\reference Vanture, D.V., Wallerstein, G., Brown, J.A. 1994, PASP, 106, 835
\reference Woolley, R., 1966, Royal Obs. Ann. London, No 2
\end{references}
\end{document}